# Exchange Enhancement of the Electron-Phonon Pair Interaction

D. Manske, C.T. Rieck, and D. Fay

Abteilung für Theoretische Festkörperphysik am Institut für Angewandte Physik,
Universität Hamburg, Jungiusstr. 11, 20355 Hamburg, Germany

The critical temperature of high-$T_c$ superconductors is determined, at least in part, by the electron-phonon coupling. We include the effect of an exchange interaction between the electrons and calculate the renormalization of the bare phonon frequencies and the electron-phonon verticies in a random phase approximation and obtain a strongly enhanced attractive phonon-induced electron-electron interaction. Using Fast Fourier Transform techniques, the weak-coupling selfconsistency equation for the order parameter is solved in the 2D first Brillouin zone for the Emery tight-binding band with different band fillings. The enhancement of $T_c$ arises primarily from the softening of the phonon frequencies rather than the vertex renormalization.

## 1. INTRODUCTION

Although it is widely believed that the phonon mechanism alone cannot yield $T_c$ of 90K or more, the role of phonons in the high-$T_c$ superconductors is still a subject of debate. Even if electronic correlations are primarily responsible for the superconductivity, the phonon mediated part of the pair interaction may still make a significant contribution and it is important to estimate the effect of electronic correlations on the phonons[1]. Since strong antiferromagnetic (AF) spin correlations are known to be present in the cuprate systems, we consider a simple two dimensional RPA-type model for the exchange-enhanced electron system, and calculate the effect of the AF spin fluctuations on the phonon frequencies and the electron-phonon interaction vertex. We then use the resulting renormalized phonon mediated interaction in the gap equation with a 2D tight-binding-structure.

## 2. RENORMALIZATION OF THE PHONONS

We employ the usual model Hamiltonian consisting of bare electrons interacting through a 2D Coulomb interaction $v(q) = 2\pi e^2/q$, bare phonons of frequency $\Omega_{\mathbf{q}}$, and a bare electron-phonon vertex with coupling constant $\alpha(\mathbf{q})$. For simplicity we take the jellium approximation. The screening of the Coulomb interaction, including an effective constant exchange interaction $U$, leads to the renormalized phonon frequency[2]:

$$\omega_{\mathbf{q}}^2 = \Omega_{\mathbf{q}}^2 - 2\Omega_{\mathbf{q}}^2 \mid \alpha(\mathbf{q}) \mid^2 \frac{\tilde{\chi}(\mathbf{q},\omega_{\mathbf{q}})}{1 + v(q)\tilde{\chi}(\mathbf{q},\omega_{\mathbf{q}})} \quad (1)$$

where

$$\tilde{\chi}(\mathbf{q},\omega) = \frac{2\chi_0(\mathbf{q},\omega)}{1 - U\chi_0(\mathbf{q},\omega)} \quad (2)$$

with the non-interacting susceptibility given by

$$\chi_0(\mathbf{q},\omega) = \sum_{\mathbf{k}} \frac{f(\epsilon_{\mathbf{k+q}}) - f(\epsilon_{\mathbf{k}})}{\omega - (\epsilon_{\mathbf{k+q}} - \epsilon_{\mathbf{k}})} \quad . \quad (3)$$

The screened electron-phonon vertex is given by

$$\bar{\alpha}(\mathbf{q},\omega) = \alpha(\mathbf{q},\omega)/\epsilon(\mathbf{q},\omega) \quad (4)$$

with the dielectric function

$$\epsilon(\mathbf{q},\omega) = 1 + (2v(q) - U)\chi_0(\mathbf{q},\omega) \quad . \quad (5)$$

In $\chi_0$ we use the 2D tight-binding Emeryband[3]

$$\epsilon_{\mathbf{k}} = 2t\left[2 - \cos(k_x a) - \cos(k_y a) - \mu\right] \quad (6)$$

with $t = 170$meV. At $\mu = 2$ (half filling), $\chi_0$ has a peak at the nesting wave vector[4], $\mathbf{Q} = (\pi,\pi)$. This gives rise to a softening of $\omega_{\mathbf{q}}$ with increasing $U$ for $\mathbf{q}$ near $\mathbf{Q}$. This effect is clearly seen in Fig.(1) where $\omega_{\mathbf{q}}$ is shown along the line $\Gamma M X \Gamma$. We have solved Eq.(1) with $\omega_{\mathbf{q}} = 0$ in $\tilde{\chi}$. This should be a good approximation near the AF transition ($U = U_c$) for $\mathbf{q} \approx \mathbf{Q}$.

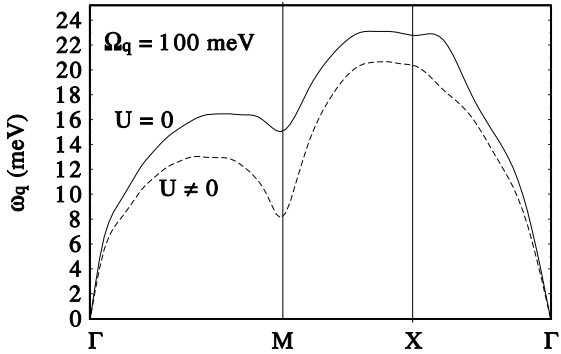

Figure 1. Renormalized phonon frequency.

It is interesting to note that, for a Hubbard model, the diagrams considered here lead to an $\omega_\mathbf{q}$ that <u>increases</u> with increasing $U$:

$$\omega_\mathbf{q}^2 = \Omega_\mathbf{q}^2 - 2\Omega_\mathbf{q} \mid \alpha(\mathbf{q}) \mid^2 \frac{2\chi_0(\mathbf{q},\omega)}{1 + U\chi_0(\mathbf{q},\omega)} \quad .$$

It is however not clear whether one double counts the screening diagrams in this case.

## 3. SUPERCONDUCTIVITY

$T_c$ is obtained by solving the linearized version of the s-wave weak coupling gap equation in the first Brillouin zone using Eq.(6):

$$\Delta(\mathbf{k}) = \int \frac{d^2\mathbf{k}}{(2\pi)^2} V(\mathbf{k} - \mathbf{k}') \frac{\tanh(E(\mathbf{k}')/2T)}{2E(\mathbf{k}')} \Delta(\mathbf{k}')$$

with

$$E(\mathbf{k}) = \sqrt{\epsilon_\mathbf{k}^2 + \Delta^2(\mathbf{k})}$$

We take for the phonon mediated electron-electron interaction in weak coupling

$$V(\mathbf{q}) = - \mid \alpha(\mathbf{q}) \mid^2 D(\mathbf{q},\omega = 0) \quad ,$$

where $D$ is the renormalized phonon propagator. The results for $T_c$ as a function of the exchange parameter $U$ are shown in Fig.2 for various band fillings. The enhancement of $T_c$ arises almost entirely from the softening of the phonon; the small electronic vertex correction seems to be typical of 2D AF spin fluctuation systems[5].

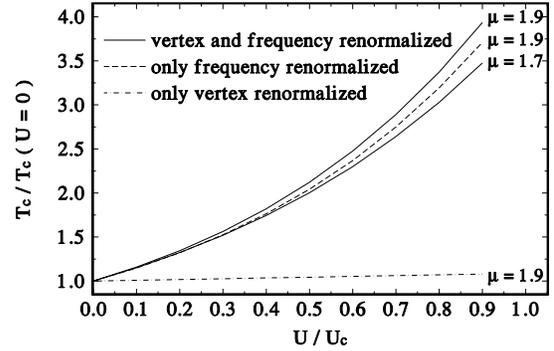

Figure 2. $T_c$ including exchange effects.

## 4. CONCLUSIONS

We conclude that the phonon-softening effect of the 2D AF spin fluctuations can considerably raise $T_c$. In order to make reliable predictions for the oxide superconductors it seems however important to do a strong coupling Eliashberg calculation since changes in the phonon coupling function $\alpha^2 F(\omega)$ can have subtle effects on $T_c$ [6]. Also one should go beyond the jellium model and consider optical phonons. We have not found any solutions of the gap equation for d-wave symmetry. If strong AF spin fluctuations lead to d-wave pairing, the corrections to the phonon-mediated part of the interaction considered here should still be relevant.